\documentclass[traditabstract]{aa}
\usepackage{natbib}
\usepackage{txfonts}
\usepackage{graphicx}
\titlerunning{The occurrence of classical Cepheids in binary systems}
\authorrunning{Neilson et al.}

\begin{document}


\title{The occurrence of classical Cepheids in binary systems}

\author{Hilding R. Neilson\inst{1,2} \and Fabian~R.~N.~Schneider \inst{3} \and Robert G.~Izzard\inst{3} \and  Nancy R.~Evans\inst{4}  \and Norbert Langer\inst{3}}
\institute{
   Department of Astronomy and Astrophysics, University of Toronto, 50 St.~George Street, Toronto, ON, Canada M5S3H4 \\
      \email{neilson@astro.utoronto.ca} 
   \and
    Department of Physics and Astronomy, East Tennessee State University, Box 70652, Johnson City, Tennessee 37614, USA 
       \and
     Argelander-Institut f\"{u}r Astronomie, Universit\"{a}t Bonn, Auf dem H\"{u}gel 71, D-53121 Bonn, Germany 
     \and
  Smithsonian Astrophysical Observations, MS 4, 60 Garden Street, Cambridge, MA 02138, USA
  }

\date{}
\abstract{
Classical Cepheids, like binary stars, are laboratories for stellar evolution and Cepheids in binary systems are especially powerful ones.  About one-third of Galactic Cepheids are known to have companions and Cepheids in eclipsing binary systems have recently been discovered in the Large Magellanic Cloud.  However, there are no known Galactic binary Cepheids with  orbital periods less than one year.  We compute population synthesis models of binary Cepheids to compare to the observed period and eccentricity distributions of Galactic Cepheids as well as to the number of observed eclipsing binary Cepheids in the LMC.   We find that our population synthesis models are consistent with observed binary properties of Cepheids.
Furthermore, we show that binary interaction on the red giant branch prevents  some red giant stars from becoming classical Cepheids.  Such interactions suggest that the binary fraction of Cepheids should be significantly less than that of their main-sequence progenitors, and that almost all binary Cepheids have orbital periods longer than one year. If the Galactic Cepheid spectroscopic binary fraction is about 35\%, then the spectroscopic binary fraction of their intermediate mass main sequence progenitors is about 40 -- 45\%.
}

\keywords{stars:binary / stars: variable: Cepheids}
\maketitle

\section{Introduction}
Classical Cepheids are intermediate-mass, evolved stars which have been observed for centuries ever since Cepheid variability was discovered by John Goodricke in 1784 \citep{Good1785}.  The source of the variability was a long-standing mystery, one suggestion was that Cepheids are eclipsing binary stars. However, the first distance measurements to nearby Cepheids implied that a Cepheid light curve could only be consistent with an eclipsing binary if the orbital separation was less than the stellar radius, negating the binary hypothesis.  While disproved, the binary hypothesis persisted until the 1920s \citep{Jean19xx}, even though \cite{Eddington1918} demonstrated that Cepheid variability is due to radial pulsation.

It is radial pulsation that makes Cepheid variable stars important for stellar physics, extragalactic astronomy and cosmology.  \cite{Leavitt1908} discovered the Cepheid period-luminosity relation (Leavitt Law) from observations of Large Magellanic Cloud Cepheids, which was applied to extragalactic observations to measure the Hubble Constant \citep{Lemaitre1927, Hubble1929}, thus founding modern cosmology. 

 Measurements of Cepheid properties such as mass, radius and luminosity constrain the detailed physics of stellar evolution and pulsation.  For instance, theories of  helium fusion \citep{Morel2010} and mass loss \citep{Neilson2012b, Neilson2012c} have been constrained by measurements of period change, while mass measurements constrain theories of convective core overshooting \citep{Keller2006, Neilson2011}.  These results have implications for late-stage stellar evolution and supernovae, as well as constraining the main sequence evolution of their progenitors.

While their variability is not a result of binary eclipses, Cepheids with binary companions are also of interest for understanding stellar physics.  Observing eclipsing binary Cepheids provides a direct measure of stellar masses that can be compared to  predictions from stellar evolution and stellar pulsation models and independently constrains the long-standing mass discrepancy \citep{Cox1980, Keller2008}.  Similarly, observing the orbital period and mass distribution of binary Cepheids constrains the binary fraction of intermediate-mass main sequence stars and the initial mass function \citep{Evans2013}.  This is of particular importance as \cite{Mayor1991} and \cite{Raghavan2010} measured the binary fraction of low-mass stars to be approximately $50\%$ while \cite{Sana2012} and \cite{Chini2012} suggested the binary fraction of high-mass stars is $70 - 90\%$. \cite{Kouwenhoven2005, Kouwenhoven2007} suggested the binary fraction of intermediate-mass stars in the cluster Scorpius OB2 is about $50\%$, leaving the binary fraction of intermediate-mass stars in the Galaxy an important and open question \citep{Kouwenhoven2008, Raghavan2010}. 

Observations are providing insight into the orbital period and eccentricity distributions of binary Cepheids in the Galaxy. \cite{Evans2005} presented a list of single-line spectroscopic binaries with measured orbital periods and eccentricities, as well as updated orbital parameters for the Cepheid Polaris \citep{Evans2008},  based on a literature review of all known Galactic Cepheids.   That binary sample is complete with the exception of small mass ratios ($< 0.2$) and long-period orbits ($>10~$yr)  and is the most complete sample of Galactic Cepheid binarity. \cite{Szabados2012} argued that the Galactic Cepheids X~Pup and XX~Sge are also spectroscopic binaries based on 75 years of observations.  \cite{Szabados2013a, Szabados2013b} analyzed spectroscopic observations to detect binary companions for a sample of Galactic Cepheids in the southern hemisphere. \cite{LiCausi2012} observed the companion of   the Cepheid X~Sgr using interferometric observations.  Similarly,   \cite{Gallenne2013, Gallenne2014} discovered, using interferometric observations, that the Cepheids V1334 Cyg and AX Cir also have companions. \cite{Evans2005} noted that about 35\% of Galactic Cepheids have at least one spectroscopic companion, and about 44\% of those have more than one companion. There are no known Galactic Cepheids with a companion with orbital period less than one year \citep{Evans2011}.

Observational studies of Cepheid duplicity in the Large Magellanic Cloud are also beginning to yield results.  Three eclipsing binaries were discovered in the OGLE-III survey of LMC Cepheids \citep{Soszynski2008} while \cite{Szabados2012b} noted there are four known spectroscopic Cepheid binaries.  \cite{Szabados2012b} also attempted to use velocity and $B$- and $V$-band amplitudes to search for new binary Cepheids. From their sample of 43 Cepheids, \cite{Szabados2012b} detected seven additional binary Cepheids.  They suggested that the binary fraction of Cepheids in the LMC may be less than that of the Milky Way, but the result still suffers from significant observational bias because of their method.

The purpose of this work is to explore Cepheid duplicity from the perspective of theoretical binary stellar evolution models and to compare predicted Cepheid binary properties with those observed in the Galaxy and LMC. As such we compute the probable minimum orbital period for Galactic Cepheids from the modeled orbital period distribution.  We also predict the number of Cepheids in eclipsing binary systems that would be detectable in the OGLE-III survey to compare with the observed number.  This work is a first step in understanding the role of binarity in Cepheid stellar evolution.  We explore the distribution of binary systems containing a Classical Cepheid as the primary component from population synthesis calculations.  We discuss the synthesis code and describe our assumptions in Section 2.  In Section 3, we present our results for Galactic and Large Magellanic Cloud metallicities.  We compare these results to the observed binary fraction in Section 4 and discuss the role of tidal evolution and mass transfer on the evolution of stars from the red giant branch to the blue loop. We conclude in Section 5.

\section{Binary population synthesis}
We use the single and binary star population nucleosynthesis code of \cite{Izzard2004, Izzard2006, Izzard2009} which is based on the rapid binary evolution code of \cite{Hurley2002}.  Formulae fitted to detailed single star models with convective core overshooting describe stellar evolution across the whole Hertzsprung-Russell diagram \citep{Pols1998, Hurley2000}. Stellar wind mass loss is given by \cite{Nieuwenhuijzen1990} for stars with luminosities $L \ge 4000~L_{\sun}$ and is also a function of metallicity \citep{Kudritzki1989}.  For giant branch evolution, we apply the \cite{Kudritzki1978} mass-loss prescription with a wind modification factor $\eta = 0.5$ \citep{Hurley2000}.  Wind mass loss during other evolutionary stages is insignificant for  Cepheid stars. Tidal evolution, however, is important because it synchronizes the orbital period and the spins of stars, and circularizes the orbit.  For stars with convective envelopes we use the equilibrium tidal model of \cite{Hut1981}, while for stars with radiative envelopes, we employ the dynamical tide models of \cite{Zahn1975, Zahn1977} based on the  \cite{Hurley2002} implementation.

We set the mass of the primary star to be $M_1 = 3$ -- $15~M_\odot$ and the mass ratio to be $q \equiv M_2/M_1= 0.1$ -- $1$, where $M_1$ is the primary mass and $M_2$ is the secondary mass.  The initial separation range is $a = 20$ -- $10^5~R_\odot$ and the eccentricity varies from zero to unity. We assume a flat distribution for the eccentricity, a uniform distribution for the logarithm of the initial separation and a \cite{Kroupa1993} initial mass function. Furthermore, we assume a constant star formation rate.

We label a star as a Cepheid when one of the stars evolves onto the Cepheid instability strip as determined by \cite{Bono2000}.  This includes core helium-burning stars, and stars evolving across the Hertzsprung gap.  However the latter population is expected to be small because their evolutionary timescale  is much shorter than for later crossings of the instability strip.  These first crossing Cepheids contribute less than a few percent to the total Cepheid population. We also count binary systems in which the secondary star is a Cepheid. If a Cepheid is the secondary then the primary is initially more massive, hence is more likely to have evolved into a white dwarf star or exploded as a supernova. Such a Cepheid binary system will be rare and difficult to observe because the relic companion will be much dimmer than the Cepheid.  However, in rare cases the initially more massive primary star will accrete onto the secondary during main sequence evolution, and the initial primary may not evolve to be a white dwarf or neutron star before the other star becomes a Cepheid.  These systems are also counted, but not considered in Sect.~3.

\section{Shortest orbital period Cepheid binary}
The two phenomena that are most important for determining the number of Cepheid binary systems are: 1) the minimum separation at which Roche-lobe overflow starts while evolving along the red giant branch and 2) the initial binary distribution of intermediate-mass stars.  The most probable minimum orbital period for a system with a Cepheid primary star will be dictated by the Roche-lobe separation of the binary system, i.e., the radius of the Cepheid when it was a red giant star, while tides play a role also.  We use the term `most probable minimum orbital period' because we assume that the Cepheid component of the binary is evolving along its blue loop, which occurs after the red giant stage, i.e., most likely to be observed.  However, Cepheid pulsation also occurs when a star crosses the Hertzsprung gap after main sequence evolution, and a binary system at this stage can have a much shorter orbital separation without interacting.  We account for this stage later, but as this evolutionary stage has a negligible life time relative to a star evolving on the blue loop, these systems will be rare.

Roche-lobe overflow and tidal interaction both depend on the ratio of the stellar radius, $R$,  to the orbital separation, $a$.  Tidal  circularization and synchronization timescales are proportional to $(R/a)^{-8} $ and $(R/a)^{-6}$, respectively \citep{Zahn1977, Hut1981}.  The primary undergoes mass transfer when its stellar radius is greater than the Roche-lobe radius  \citep{Paczynski1971, Eggleton1983}.  For $q = 1$, mass transfer starts when the radius is about one-half the orbital separation.  

Mass transfer from a red giant star is unstable in our models, hence leads to common envelope evolution even though mass transfer depends on the mass ratio and angular momentum  of the binary system.   Unstable mass transfer occurs when the mass donor cannot readjust to hydrostatic equilibrium on a dynamical time scale. Once common envelope evolution ends, the primary is left as a helium star which, because it has negligible envelope material, is never a classical Cepheid.  But, even if mass transfer is stable, the primary star's envelope will still be stripped.  Therefore, we estimate the minimum orbital period for a Cepheid evolving on the blue loop as the primary star.

Following \cite{Evans2011}, we assume that a typical binary system with a Cepheid primary has a primary mass $M_1 = 5~M_\odot$ and secondary mass $M_2 = 2~M_\odot$.  We compute a $5~M_\odot$ stellar evolution model using the \cite{Yoon2005} stellar evolution code assuming mass loss from \cite{Kudritzki1989} for hot stars and \cite{dejager1989} for cool stars, convective core overshooting $\alpha_c = 0.2$ \citep[for details see][]{Neilson2012b, Neilson2012c} and the \cite{Grevesse1998} solar metallicity, $Z = 0.02$.  We do not include rotation in the models. \cite{Anderson2014} and \cite{Neilson2014b} both suggest that Cepheids cannot evolve from rapidly-rotating main sequence progenitor stars  with a equatorial rotational velocity $v_{\rm{rot}} \ge 0.4v_{\rm{crit}}$. The stellar radius at the tip of the red giant branch is $R= 125~R_\odot$.   
\begin{figure*}[t]
\includegraphics[width=0.5\textwidth]{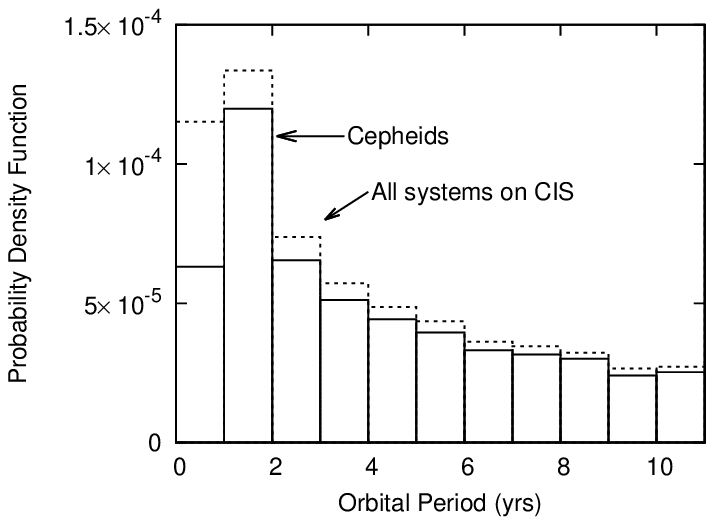}\includegraphics[width=0.5\textwidth]{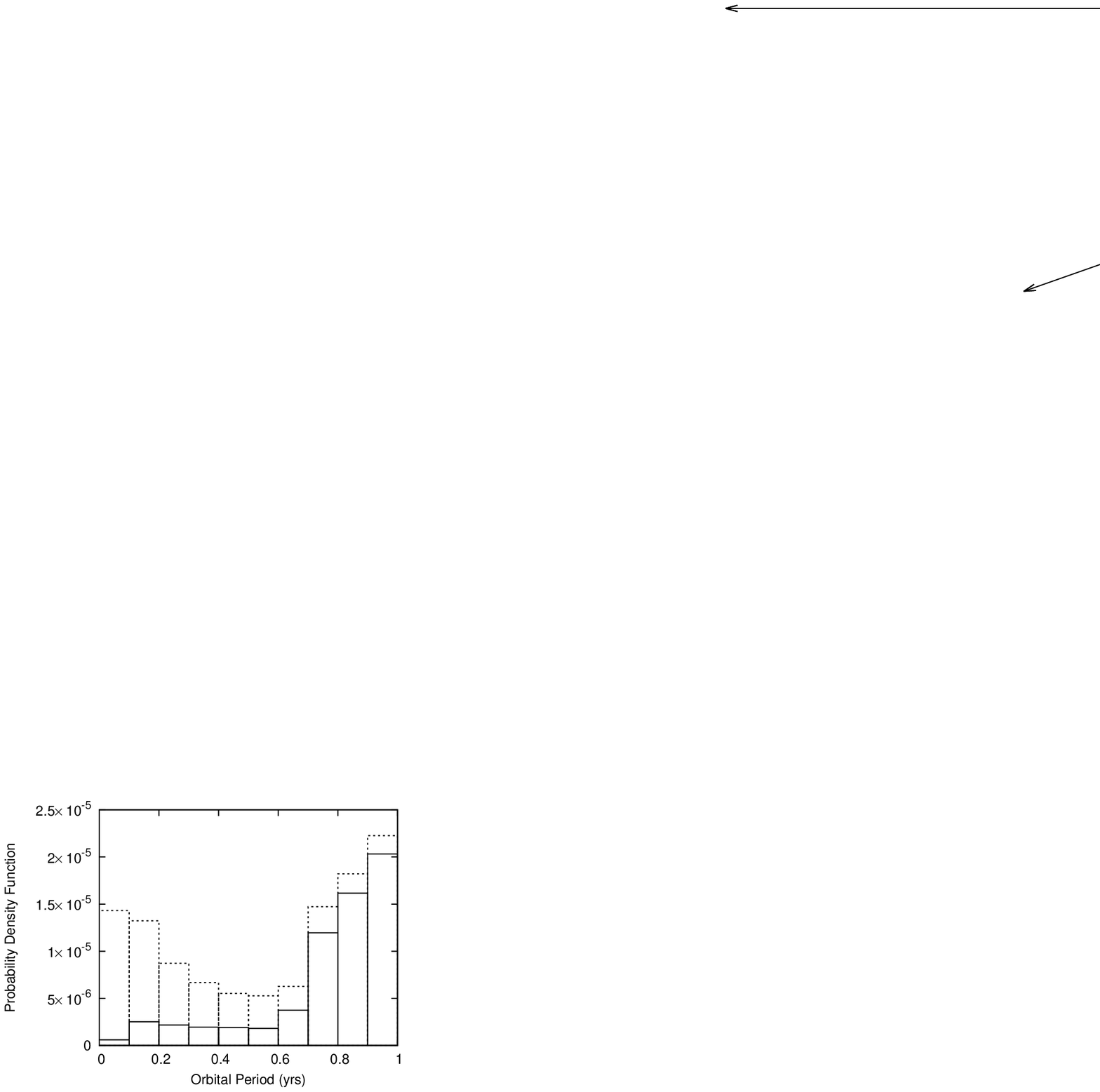}
\caption{The orbital period distribution of stars residing inside the Cepheid instability strip at solar metallicity, $Z = 0.02$, according to our population synthesis models. (Left) Binary systems for the entire period range, systems with an orbital period longer than 11 yrs are not plotted.  (Right) Systems with orbital periods shorter than $1$~yr.  The dotted histogram represents all systems evolving across the instability strip while the solid histogram represents those systems identified as Cepheids, i.e., with no significant mass transfer from the primary to secondary.}
\label{f1}
\end{figure*}

The Roche-lobe radius of a system with mass ratio $q = 0.4$ is $r_L \approx 
0.4a$ \citep{Eggleton1983} and  we assume that $R < r_L$ else the star begins mass transfer.   The minimum orbital separation that allows a red giant to evolve to a Cepheid is  $a_{\rm{min}} = 312~R_\odot$ or an orbital period $P_{\rm{orb}} > 240~$days or 0.66~years, if we ignore tides.

Tides are an additional constraint on the minimum orbital period for the binary system. We compute the change of orbital separation due to angular momentum transfer from the orbit to the rotation of the $R = 125~R_{\sun}$ red giant star.  If tides are important during red giant evolution, then tides will reduce the orbital separation and can lengthen the expected minimum orbit.  We assume the red giant is non-rotating and becomes synchronized with the minimum orbital period at which Roche lobe overflow begins, assuming conservation of angular momentum. This is an  extreme estimate because if the primary star is already rotating then less angular momentum is transferred by tides. The change of rotational angular momentum of the red giant star is,
\begin{equation}
\Delta J \approx I\Delta \omega = I \frac{2\pi}{P_{\rm{RL}} },
\end{equation}
while the change of orbital angular momentum, for $e = 0$,  is 
\begin{equation}
\Delta J = \left(\frac{GM_1^2M_2^2}{M_1+M_2}\right)^{1/2} (a^{1/2}_{\rm{min}} - a^{1/2}_{\rm{RL}}).
\end{equation}
In Eq.~1, $I = kMR^2$ is the moment of inertia, $k \approx 0.2$ depends on the density structure of the star and $P_{\rm{RL}}$ is the orbital period at which Roche lobe overflow begins.  In Eq.~2, $a_{\rm{RL}}$ is the orbital separation at which the stars begin Roche lobe overflow and $a_{\rm{min}}$ is the minimum initial  orbital separation for which tides decrease the orbital separation to $a_{\rm{RL}}$. Combining Eqs.~1 and 2, we find,
\begin{equation}\label{et}
a_{\rm{min}} = \left[a_{\rm{RL}}^{1/2} + I\left(\frac{2\pi}{P_{\rm{RL}}}\right)\left(\frac{M_1+M_2}{GM_1^2M_2^2}\right)^{1/2}\right]^2.
\end{equation}
Assuming $R = 125~R_{\sun}$, $M_1 = 5~M_{\sun}$, $M_2 = 2~M_{\sun}$, $a_{\rm{RL}}= 312~R_{\sun}$, $P_{\rm{RL}} = 0.66$~years and that $k \sim 0.2$ then the minimum orbital separation, during red giant evolution,  is $a_{\rm{min}} = 388~R_{\sun}$ and $P_{\rm{min}} = 333~$days $= 0.91~$years, similar to the minimum observed period for Galactic Cepheids.  This minimum period is not an absolute limit, there may exist Cepheid binary systems with orbital periods between 240 and 333 days, but they become increasingly rare for shorter orbital periods.  Because of the non-linear nature of Eq.~\ref{et}, the change of orbital separation decreases with increasing initial separation.  Therefore, we expect a Gaussian-like period distribution about the characteristic time scale $P = 333~$days based on interactions during the red giant stage of evolution.

 Binary systems with initial separations shorter than $388~R_{\sun}$  will interact and undergo mass transfer during the red giant evolution.  Systems with longer separations will not interact significantly, making this a reasonable estimate for the most probable minimum separation.  In principle, if the star is rotating with non-zero velocity then the minimum separation could be smaller, however, Cepheids are slow rotators, $\le 10~$km~s$^{-1}$ \citep{Bersier1996}, so rotation is not likely to play a significant role during Cepheid and red giant evolution. Similarly, mass loss rates are not large enough to significantly widen the binary orbit.  

It is remarkable that our predicted most probable minimum orbital period agrees so well with the observed distribution of Galactic binary Cepheids and suggests that tidal evolution is important for binary evolution. Of particular interest is the recent observation that X Sgr is a binary system with an orbital separation of $385~R_\odot$ \citep{LiCausi2012}, coincident with the results here.  Our analysis is limited to the mass ratio $q = 0.4$ and $e = 0$, different mass ratios and non-zero eccentricities lead to different minimum orbital periods.   If the mass ratio is greater than 0.4 then $a_{RL}$ increases, however assuming  $M_1$ is still $5~M_\odot$ then decreasing the mass ratio decreases the orbital angular momentum and the ratio $(M_1 + M_2)/(M_1^2M_2^2)$ increases.  Because that term is divided by the orbit period at the moment Roche lobe overflow begins, then the changes due to different mass ratios will mostly cancel and we will be left with a similar minimum orbital separation.  If, instead, we assume a non-zero eccentricity then $a_{RL}$ is unchanged but the orbital angular momentum changes.  Hence,  $a_{\rm{min}}$ will increase.  Also, as discussed above, there may also exist a small number of close binary Cepheids in which the Cepheid is evolving along the first crossing of the Cepheid instability strip and is not yet a red giant star. We will show that this scenario occurs in about 3 - 5\% of binary Cepheids.

\section{Cepheid binary period distribution}
Our estimate for the binary Cepheid with the shortest orbital period depends on the assumed masses of the two stars, yet Classical Cepheids range in mass from about $4~M_\odot$ to $15~M_\odot$. We study the shortest orbital period and number of Cepheid binary systems for a population of stars evolving on the Cepheid instability strip by computing population synthesis models at $Z=0.02$ and $Z=0.008$. This sample includes stars evolving through the blue loop and the first crossing of the Hertzsprung gap.

\subsection{Galactic Cepheid distribution}
In Fig.~\ref{f1} we plot the period distribution of binary systems in which the primary, star 1, or the secondary, star 2,  resides within the Cepheid instability strip as represented by the dotted histogram.  We find that about 8\% of these systems have an orbital period 
shorter than one year and a significant fraction of those systems have an orbital period shorter than $0.1$~years. However, the vast majority of them have undergone mass transfer. These systems are not expected to produce Classical Cepheids because the primary star loses most of its envelope. When we ignore those systems in which the primary loses more than $10\%$ of its initial mass,  the orbital period distribution (the solid histogram, Fig.~\ref{f1}) still contains a few systems ($\approx5\%$) with orbital period shorter than one year and  an even smaller number of systems with orbital period shorter than 0.5 year.   One constraint also includes mass lost in a wind, but for these stars total mass lost before the Cepheid blue loop is less than $1\%$ of the total mass.  Also,  about 3\% of binary Cepheids have an orbital period shorter than the minimum predicted in Sect.~3, though this number depends weakly on the maximum binary separation assumed in our model.   These short orbital period systems correspond to Cepheids evolving along the Hertzsprung gap and those Cepheids that are secondary components with a more evolved companion.  The latter scenario occurs when the primary donates material to the secondary, so much so that the mass ratio inverts and the secondary becomes more massive than the primary.  This causes the binary orbit to widen and what was the secondary evolves to become a Cepheid.

\subsection{Comparison to observed binaries}
\begin{figure}[t]
\includegraphics[width=0.5\textwidth]{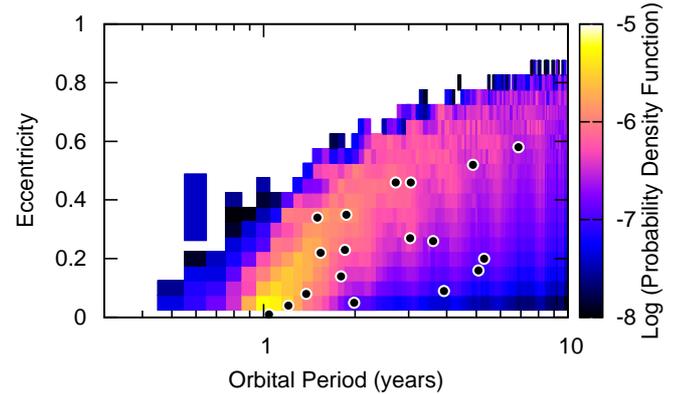}
\caption{Comparison of the orbital period and eccentricity distributions from our population synthesis models and properties of observed Galactic Cepheid binary systems \citep{Evans2005}, denoted with dots. The colour scale represents the total probability that a binary of given properties exists assuming constant star formation.}
\label{fig:map}
\end{figure}

For further comparison, we plot the probability map of the orbital period and eccentricity for Galactic Cepheids in Fig.~\ref{fig:map}, along with the periods and eccentricities for a sample of Galactic Cepheid binaries from \cite{Evans2005}.  Many of these spectroscopic binaries were detected from ultraviolet observations. All Cepheid binaries in the sample are spectroscopic binaries with the exception of Polaris which is an astrometric binary \citep{Kamper1996, Evans2008}.  One-half of the Cepheids in the sample have orbital periods between one and two years and most have eccentricities less than $0.5$. There may be an observational bias towards short-orbital period Cepheids because they are more easily identified as spectroscopic binaries.

There is some agreement between our model and the distribution of Galactic Cepheid binaries.  The observed number ratio between binary Cepheids with orbital periods $< 2$ yrs and those $2<P_{\rm{orb}}<8$~yrs is 50:50.  From the computed models, the ratio is 41:59, consistent with the observed sample of 18 Cepheids.  While the orbital period distribution appears to be consistent, there are distinct differences between the observed and predicted eccentricities for binary Cepheids with orbital periods $> 2$ yrs. Observed binaries with $P_{\rm{orb}} >2$ yrs appear to have more circular orbits than our model predicts.  Specifically, there appear to be five systems with orbital periods longer than two years and small eccentricity ($<0.25$), otherwise the agreement between our model and observations is remarkable.  The difference for those five systems is likely because of the assumed initial eccentricity distribution and resolution of the models but, without a greater number of observed binary Cepheids to which we can compare, it is difficult to quantify the eccentricity discrepancy between the predictions and observations  and whether it is significant.

There is a peak in the probability distribution function near an orbital period $P_{\rm{orb}} \approx 1$~yr which is consistent with the observed binary sample \citep{Evans1995}.  This peak occurs for eccentricities ranging from $e = 0$ -- $0.6$ for orbital periods ranging from one to three years, also in agreement with the observed sample, confirming our predictions from the previous section.  However, it is unclear whether our model agrees with observations of systems with longer orbital periods.
 
\begin{figure*}[t]
\includegraphics[width=0.5\textwidth]{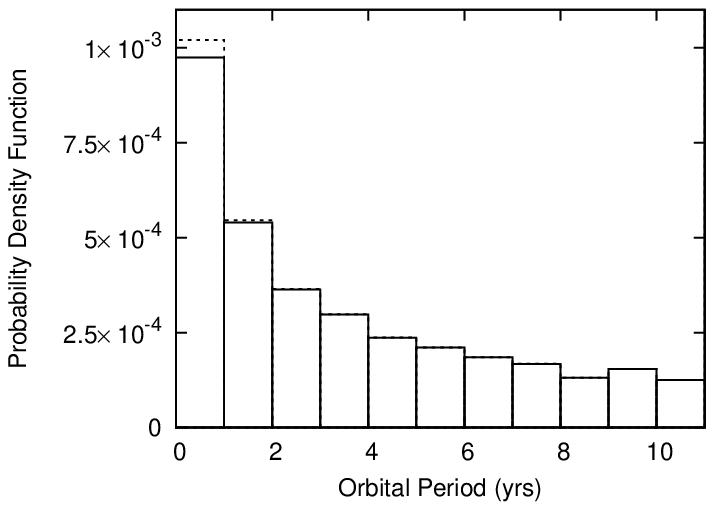}\includegraphics[width=0.5\textwidth]{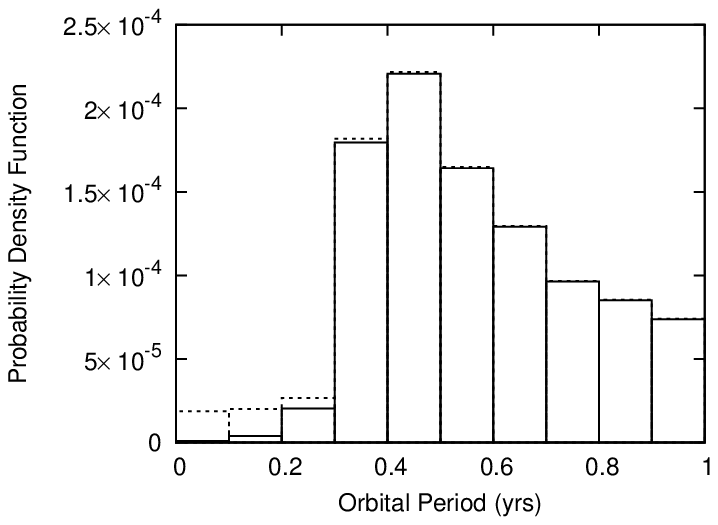}
\caption{The orbital period distribution for stars crossing the Cepheid instability strip at LMC metallicity, $Z = 0.008$. Dotted and solid lines have the same meaning as in Fig.~\ref{f1}.}
\label{f2}
\end{figure*}

\subsection{LMC Cepheid period distribution}

We also compare population synthesis models assuming metallicity $Z=0.008$ with the OGLE-III survey of classical Cepheids and  plot the predicted period distributions in Fig.~\ref{f2}.  The period distribution of LMC stars evolving along the Cepheid instability strip differs from Galactic stars.  The LMC binary systems do not appear to interact as often and fewer stars lose a significant fraction of their initial mass.  There are more Cepheid binaries with orbital period shorter than one year relative to all systems including mass transfer.  By considering those binary Cepheids with orbital period shorter than $1$~yr, we find a peak in orbital periods at about $0.4$~yr as opposed to between one and two years for Galactic Cepheids. The radius of an LMC star at the tip of the RGB is smaller than for a star with the same mass but with Galactic metallicity suggesting that, for a given separation, $a$, tides are also weaker. Furthermore,  stellar evolution models form Cepheid blue loops at lower mass at smaller metallicity \citep{Pols1998,Bono2000b}, implying that Cepheid binary systems can evolve to have smaller separations.  Binary separations must also be shorter to begin Roche-lobe overflow.

\cite{Soszynski2008} presented OGLE-III survey observations of 1848 fundamental-mode, 1228 first-overtone and  285 other types of Classical Cepheids in $V$ and $I$-band wavelengths for seven years.  Three confirmed eclipsing binary systems with a fundamental-mode Cepheid component were observed, with subsequent analysis by \cite{Piet2011a, Piet2011b} for two of them.  Perhaps the third system, OGLE-LMC-CEP1718, is particularly unusual, being composed of two apparently equal mass Cepheids pulsating in the first overtone, but with different pulsation periods \citep{Gieren2014}. Four more Cepheids are suspected to be in eclipsing binary systems.  We compare to the observed LMC Cepheid eclipsing binary fraction of $3/3361 = 0.089\%$ to $7/3361 = 0.2\%$.  We limit our analysis to the orbital-period range $< 3$ years, because only these binary systems would show multiple ($> 1$) eclipses in the seven-year OGLE-III survey.

 \cite{Evans2013} found that 55\% of Cepheid binaries with q $>$ 0.4 are spectroscopic binaries.  Because the observed percentage of spectroscopic binaries over all mass ratios is about 35\%, then we suggest that a total binary fraction of Cepheids is $35\%/0.55 =  64\%$. The spectroscopic binary fraction is compiled from a broad sample of Galactic Cepheids and may suffer some biases, but is the most complete sample available. \cite{Szabados2012b} found a smaller binary fraction for LMC Cepheids but admit an observational bias. As such, we assume that Cepheids in the LMC have a binary fraction between $50\%$ and $64\%$.  Therefore, of the 3361 fundamental-mode Cepheids observed in the OGLE-III survey, between 1680 and 2185 Cepheids have companions.  The fraction of LMC Cepheid binaries with orbital periods shorter than $3$ yr is
about $ 27\%$ in our model, suggesting 454 - 590 Cepheids observed in the LMC have companions with orbital period less than $3$ yr.  

To convert the expected number of binary Cepheids to the number of eclipsing binary Cepheids, we estimate the number of systems with an inclination that gives eclipses. The two LMC binary Cepheids that have been analyzed have orbital inclinations $i = 87^{\circ}$ and $90^{\circ}$.  If we consider a typical binary system with a $5~M_\odot$ Cepheid and a $2~M_\odot$ companion and orbital period of $0.4$~yr and if we assume the limiting inclination to produce eclipses is $i=87^{\circ}$  then same binary with a $3$~yr orbit has a limiting inclination of $i =  89.2 ^{\circ}$. 

An inclination range between $87^{\circ}$ and $89.2^{\circ}$ suggests an eclipsing binary fraction of $0.9\%$ and $3.3\%$, i.e. between $4$ and $20$ Cepheids observed in the OGLE-III survey should be in eclipsing binary systems. This is consistent with the number presented by \cite{Soszynski2008} of three to seven Cepheid eclipsing binaries.

\section{Discussion}
We compute population synthesis models of Cepheid binary systems for both Galactic and LMC metallicities.  Our synthesis models of Galactic Cepheids predict a minimum orbital period of about 300~days (0.82~years), consistent with $P = 381$~days for $Z$ Lac \citep{Evans1995}.  The shortest orbital period for a binary Cepheid evolving along the blue loop depends, primarily, on the Roche lobe separation and the efficiency of tides when the Cepheid was evolving previously as a red giant when its radius was greatest. We also find agreement between our predicted number of LMC eclipsing binary Cepheids and those discovered in the OGLE-III survey.

We predict a period distribution that agrees with observations, but we argue that a significant fraction of binary systems with stars evolving on the instability strip do not form Cepheids because of mass transfer from the star that would be a Cepheid.  This suggests that the binary fraction of Cepheids is smaller than the binary fraction of main sequence B-type stars that are Cepheid progenitors. \cite{Evans2005} suggest the spectroscopic binary fraction of Cepheids is 35\% while other works suggest the spectroscopic binary fraction of main sequence B-type stars range from $30\%$ to $40\%$ for Galactic stars \citep{Chini2012} and $50\%$ for the cluster Sco OB2 \citep{Kouwenhoven2005, Kouwenhoven2008}.  It should be noted that \cite{Chini2012} considered only binaries with mass ratios $> 0.2$. Our results suggest a binary fraction for Cepheids should be much smaller than that for main sequence B-type stars.  If we assume that 35\% of Galactic Cepheids are in spectroscopic binary systems,  we infer from our models that the spectroscopic binary fraction of their intermediate-mass main-sequence progenitors is about 40 -- 45\%.  This is based on computing the total probability fractions of Cepheids relative to all systems shown in Fig.~\ref{f1}.  Furthermore, this suggests a total binary fraction of intermediate-mass main sequence stars to be about 73 -- 82\%, consistent with \cite{Sana2012} and \cite{Chini2012}. We note that \cite{Evans2013} estimate a total Cepheid binary fraction to be about 64\%, consistent with our results. 

These results, though, depend on our understanding of the structure of the Cepheid blue loop. For instance, the minimum orbital period of about one year depends on the assumed masses and radii of the stars.  If smaller-mass stars evolve along the Cepheid blue loop, then the minimum separation of stars in a binary system can be shorter.  The observed period distribution of binary Cepheids potentially provides insight into the structure of blue loops even though blue loop structures are sensitive to convective core overshooting \citep{Bono2000}, metallicity \citep{Keller2008},  mass loss \citep{Neilson2012a, Neilson2012b} and other model physics. While promising, we require measurements of many more binary Cepheids to draw conclusions regarding the blue loop.
  
While our results are consistent with the number of LMC eclipsing binary Cepheids, it is difficult to directly predict the evolutionary path of OGLE-LMC-CEP1812 \citep{Piet2011b}.  The system consists of a Cepheid primary and a red giant companion in a 552~day orbit but single star models suggest that the companion must be 100~Myr older than the Cepheid \citep{Piet2011b}. The authors suggest that the binary system formed by a stellar capture. Another possibility is the system evolved from a hierarchical triple system. This particular binary system is unexplained by our binary evolution models.  The binary system OGLE-LMC-CEP1718 with two first-overtone Cepheids of equal mass, but different pulsation periods, also appears strange and may require some small amount of mass transfer to explain the systems properties \citep{Gieren2014}. However, the system OGLE-LMC-CEP0227 \citep{Piet2011a} is robustly modelled \citep{Cassisi2011, Neilson2012a,Prada2012}, suggesting that not all of the LMC eclipsing binary systems are strange. Therefore, we need to consider alternative stellar evolution scenarios to explain the existence of the system OGLE-LMC-CEP1812 that will be discussed in a forthcoming article.
 
In summary, we find that the combination tides and Roche-lobe overflow prevent the formation of short-orbital period binary systems containing a Cepheid and a companion that would be detected in various surveys such as \cite{Evans2005, Szabados2012}.  However,  a small fraction of binary systems may interact and the primary accretes mass onto the secondary star.  The primary loses its envelope and will evolve to become white dwarf stage without ever being a Cepheid, while the mass gainer will be rejuvenated and eventually evolve as Cepheid with a white dwarf companion. that type of binary system is  difficult to detect, even with precision measurements of Cepheid light curves \citep{Derekas2012, Neilson2014}.  We will explore further in future work.

\acknowledgements
HRN and RGI thank the Alexander von Humboldt Foundation and HRN thanks the National Science Foundation (AST-0807664) for funding. FRNS acknowledges funding by BCGS (DFG). NRE acknowledges funding from the Chandra X-ray Center NASA Contract NAS8-03060. We also thank the referee for helpful comments that have improved this work.

\bibliographystyle{aa} 

\bibliography{ceph_bin}

\end{document}